# Cohesion of regolith: Measurements of meteorite powders

Yuuya Nagaashi, Takanobu Aoki, and Akiko M. Nakamura

Graduate School of Science, Kobe University, 1-1 Rokkodaicho, Nada-ku, Kobe 657-8501, Japan
y.nagaashi@stu.kobe-u.ac.jp

## Abstract

The cohesion of particles has a significant effect on the properties of small bodies. In this study, we measured in open air, the cohesive forces of tens of micron-sized irregularly shaped meteorite, silica sand, glass powder, and spherical glass particles, using a centrifugal method. In addition, we estimated the amount of water vapor adsorbed on the particles under the measurement conditions. The measured cohesive forces of the meteorite particles are tens of times smaller than those of an ideally spherical silica particle and correspond to the submicron-scale effective (or equivalent) curvature radius of the particle surface. Moreover, based on the estimated amount of water vapor adsorbed on the particles, we expect the cohesive forces of the particles in airless bodies to be approximately 10 times larger than those measured in open air. Based on the measurement results, we estimate that the cohesive forces of the particles on asteroids are typically in the sub-micro-Newton range, and that the particles on fast-rotating asteroids are tens of microns in size.

# 1. Introduction

The cohesion and shape of the particles constituting small bodies have a significant effect on the physical properties of the bodies, and are the keys to understanding the evolutional processes of the bodies. Cohesive forces between similar types of particles or adhesive forces between different types of particles, driven by interactions such as van der Waals, can be considerable compared with the gravitational forces acting on the bodies (Scheeres et al., 2010). We use the term "cohesive" for simplicity hereinafter even though the term "adhesive" would be more appropriate. The interparticle cohesive forces can maintain the porosity of the regolith layer because cohesive forces hinder the reconfiguration of the particles caused by the small gravitational force on the regolith layer on small bodies (Kiuchi and Nakamura, 2014). The porosity of the regolith layers affects the thermal conductivity of (Gundlach and Blum, 2013, Sakatani et al., 2016) and the formation of impact craters on (Housen et al., 2018) the surfaces of small bodies. Moreover, the additional strength induced in rubble piles by the cohesive forces may allow the bodies to rotate fast and gravity alone cannot explain the phenomenon (Sánchez & Scheeres, 2014; Rozitis et al., 2014). The cohesive force of the particles can also affect the deformation and failure modes of the rotating self-gravitational aggregates, that is, rubble piles (Hirabayashi et al., 2015; Sánchez & Scheeres, 2016). The cohesion of particles is an important parameter affecting the physical properties of small bodies, but also for the early stages of dust growth in protoplanetary disks. The cohesive forces of protoplanetary dust determine the critical dust aggregate size that can be grown by collisional sticking (Weidenschilling, 1980; Blum and Wurm, 2008).

The measurement of the cohesive forces of micron-sized silica spheres using atomic force microscope cantilevers at an ambient pressure of $10^2$–$10^5$ Pa and a humidity of 10%–40% confirmed that the cohesive forces increase proportionally to the particle size (Heim et al., 1999), which is consistent with the Johnson–Kendall–Roberts (JKR) theory (Johnson et al., 1971). However, the extrapolation of these measurements predicts forces orders of magnitude greater than those of 30–70 μm spherical glass beads measured in the atmosphere using atomic force microscope cantilevers (LaMarche et al., 2017) and a centrifugal method (Nagaashi et al., 2018). This discrepancy is probably due to microscopic asperities on the surface of the particles (Scheeres et al., 2010; LaMarche et al., 2017; Nagaashi et al., 2018).

Not only the asperities, but also the adsorbed water vapor reduce the measured cohesive forces of the particles, which were exposed to the Earth's atmosphere before the measurement. Direct shear tests were conducted on several soil samples simulating lunar regolith under both Earth's atmospheric and ultrahigh vacuum conditions to quantitatively evaluate the effect of lunar environmental conditions on the properties of the lunar regolith layer. The shear strength of the samples significantly increased under ultrahigh vacuum conditions (Nelson and Vey, 1968). This increase was due mainly to the reduction of the water vapor molecules adsorbed on the particles when exposed to the Earth's atmosphere. The experimental results were roughly reproduced by a model, which introduced "surface cleanliness" as a parameter to quantitatively describe the thicknesses of the gas adsorbed on the soil particles, which is determined by potential energy balance (Perko et al., 2001). However, the experimental results and the model suggest that reproducing the surface conditions of particles in airless environments is difficult by merely reducing the atmospheric pressure without heating the sample once it has been exposed to the Earth's atmosphere.

The macroscopic shape of the particles can also have a significant effect on the mechanical properties of granular systems. Irregularly shaped particles may allow a higher critical impact velocity for sticking than spherical particles (Poppe et al., 2000). Free-falling streams of irregularly shaped particles are more likely to agglomerate than those of spherical particles, which cannot be explained by the measured adhesive forces (Nagaashi et al., 2018). Moreover, the layers of irregularly shaped particles generally have higher porosity than the layers of spherical particles (Suzuki et al., 2003; Kiuchi and Nakamura, 2014). Because irregularly shaped particles have larger internal friction angles than spherical particles, the macroscopic shape and the cohesion of the particles affect the size of the impact craters that form (Uehara et al., 2003). Regolith on small bodies is considered fragments produced by impacts (Housen et al., 1979) or by thermal fatigue due to day-night temperature cycles on the surfaces of the bodies (Delbo et al., 2014). Consequently, regolith particles are in general irregularly shaped.

The ratios of the long axis $a$, intermediate axis $b$, and short axis $c$ lengths of the ellipsoidal approximation and the circularity of a particle (Appendix A) are often used to describe the macroscopic shape of the particle. The $b/a$ for Itokawa, Eros, and Ryugu boulders are ~0.68, 0.71–0.73, and ~0.68, respectively (Michikami et al., 2010, 2018, 2019a), and the $b/a$ and $c/a$ for Itokawa particles recovered by the Hayabusa mission are $0.71 \pm 0.13$ and $0.43 \pm 0.14$, respectively (Tsuchiyama et al., 2011), which are similar to those of impact fragments whose $a$:$b$:$c$ is ~$2$:$\sqrt{2}$:$1$ (Fujiwara et al., 1978). The circularity of Itokawa rocks is also similar to that of impact fragments (Aoki et al., 2014; Appendix A). Moreover, as noted recently, Itokawa particles are likely to be impact fragments in terms of crack growth (Michikami et al., 2019b).

The measurements of the cohesive forces of micron-sized spherical particles for protoplanetary dust cannot be extrapolated to those of larger irregularly shaped regolith particles because the shape, surface asperities, and composition of the spherical silica particles previously measured are different from those of regolith particles. Here, we investigated the cohesive forces of meteorite particles tens of microns in size. We first measured the axial ratio, circularity, and microscopic surface topography of the particles, and then measured the cohesive force using a centrifugal method. In addition, we measured the amount of water vapor adsorption on the particles. Finally, the cohesive strength of the regolith layer on small bodies is discussed.

## 2. Experiments
2.1. Particles

We used Murchison (CM2) and Allende (CV3) meteorites for the carbonaceous chondrite samples; Northwest Africa (NWA) 539 (LL3.5), NWA 1794 (LL5), and NWA 542 (LL6) for the ordinary chondrite samples; and Millbillillie for a eucrite sample. Carbonaceous and ordinary chondrites are composed of particles including chondrules and fine-grained (generally <5 µm) matrix silicate-rich particles, whereas eucrites are basalts (Weisberg et al., 2006). We crushed the pieces of the meteorites using an agate mortar and pestle. The meteorite particles were sorted using sieves with 38 and 75 µm mesh openings and ethanol to obtain meteorite particles with median diameters in the range of 48–65 µm. For comparison, we also prepared polydispersed spherical glass beads, and irregularly shaped glass powder (Fujikihan Co., Ltd.), and silica sand (Miyazaki Chemical Co., Ltd.) particles with median diameters of 44–48 µm. The size ranges of the glass powder, spherical glass beads, and silica sand particles were adjusted as done for the meteorite particles. The electron microscope images of the samples are shown in Figures 1a-i. The meteorite particles seem to have rougher surfaces and rounder shapes than the glass powder or silica sand particles. The distributions of the equivalent circular diameter of the samples measured from the images acquired by attaching a digital camera to an optical microscope are shown in Figure 1j. The size distributions of particles acquired by the images were consistent with those determined by a laser confocal refractometer (Nagaashi et al., 2018). The median diameters and particle densities of the samples are summarized in Table 1. Density measurements were performed on a sufficient quantity of meteorite samples (Murchison, Allende, NWA 1794, and Millbillillie) using a helium gas pycnometer (AccuPyc II 1340). The measured densities of Murchison, Allende, and Millbillillie samples, 2.94, 3.56, and 3.25 g/cm$^3$, respectively, are consistent with the grain densities of these meteorites measured in previous studies using helium ideal-gas pycnometry, which were 2.96, 3.66, and 3.2 g/cm$^3$, respectively (Macke et al., 2011a; Macke et al., 2011b). The measured density of NWA 1794, 3.45 g/cm$^3$, is within the grain density range of LL chondrites, which is 3.54 ± 0.13 g/cm$^3$ (Consolmagno et al., 2008). According to the literature, the bulk porosities of hand-sized samples of Murchison, Allende, and Millbillillie can vary over a wide range with their average values standing at 22%, 22%, and 11%, respectively (Macke et al., 2011a; Macke et al., 2011b), whereas the porosity of LL chondrites is approximately 10% (Consolmagno et al., 2008). The bulk density of meteorite particles is probably smaller than their grain density; however, it could be larger than that of hand-sized samples, depending on the type of the meteorite and the size of the void spaces (Consolmagno et al., 2008). We refer to the measured density of Murchison, Allende, and Millbillillie, as well as to that of NWA 1794 for the densities of the LL3.5 and LL6 samples, as the upper limit of the particle density hereinafter.

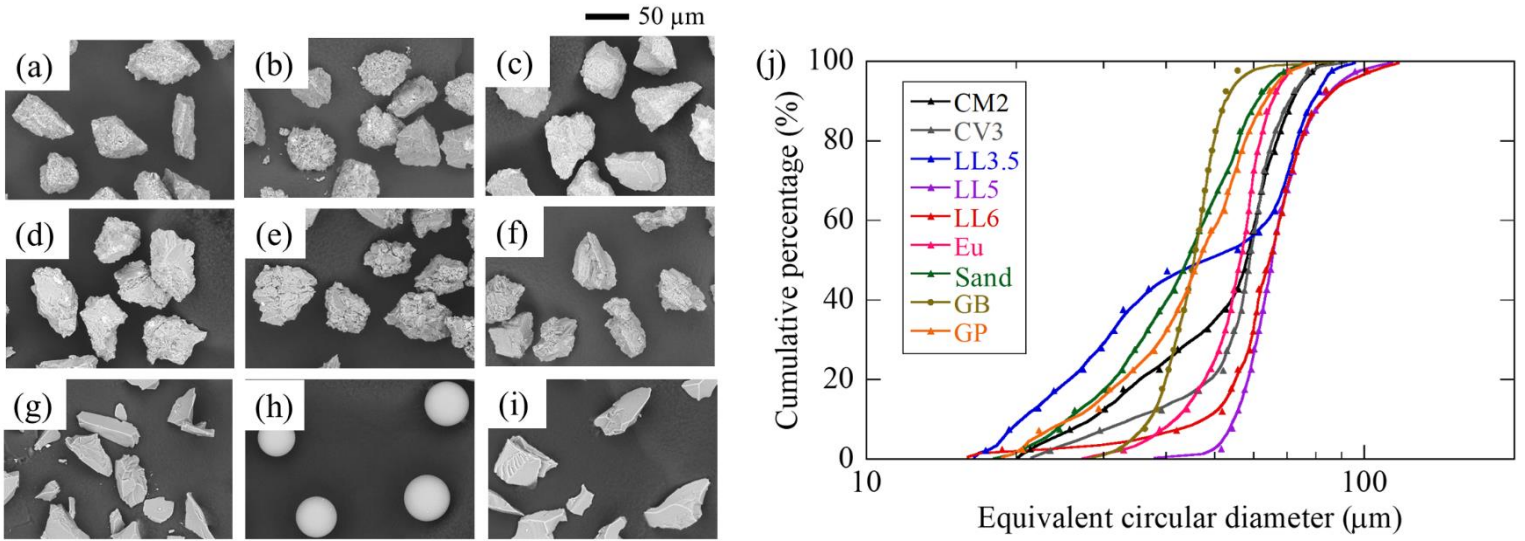

Figure 1. Scanning electron microscopy (SEM) images and size distributions. (a-i) SEM images of (a) Murchison meteorite (CM2), (b) Allende meteorite (CV3), (c) NWA 539 (LL3.5), (d) NWA 1794 (LL5), (e) NWA 542 (LL6), (f) eucrite (Eu), (g) silica sand particles (Sand), (h) glass beads (GB), and (i) glass powders (GP). (j) Particle size distributions.

Table 1. Median equivalent circular diameter, particle density, and the numbers and fraction of the samples in the cohesive force measurement.

|  | CM2 | CV3 | LL3.5 | LL5 | LL6 | Eu | Sand | GB | GP | Large CM2 | Large GP | Large Sand |
|---|---|---|---|---|---|---|---|---|---|---|---|---|
| Median Diameter (µm) | 58 | 59 | 48 | 65 | 65 | 57 | 44 | 45 | 47 | 570 | 1100 | 450 |
| Density (kg/m$^3$) | 2941.5 | 3563.7 | 3452.9 | 3452.9 | 3452.9 | 3250.9 | 2645 | 2500 | 2500 | 2941.5 | 2500 | 2645 |
| Number | 219 | 159 | 113 | 175 | 104 | 276 | 328 | 349 | 145 | 553 | 101 | 387 |
| Fraction | 0.98 | 0.99 | 0.98 | 1.0 | 0.98 | 0.99 | 0.94 | 0.97 | 0.99 | - | - | - |

2.2 Measurement of the macroscopic shapes and microscopic surface topographies of the particles

We obtained the axial ratio and circularity of the particles from the two-dimensional images of the samples (resolution: 1.3 µm/pixel) obtained using ImageJ (Schneider et al., 2012). To establish the empirical relationship between the size and the mass of irregularly shaped particles, we prepared another set of Murchison (CM2) and glass powder samples with particle sizes of 570 μm and 1100 μm, respectively, which are larger than those of the samples prepared for measuring the cohesive force. We measured the projected areas and long axes of the individual particles and the total masses of 553 Murchison particles and 101 glass powders. We also measured the axial ratio of 450 μm silica sand particles.

A confocal laser scanning microscope (LEXT OLS3100) was used to obtain particle-wide surface profiles for one or two particles of the samples with horizontal and vertical resolutions of 0.125 μm and 0.01 μm, respectively. We also measured the surface profile of the glass slide used for measuring the cohesive force (Section 2.3).

2.3 Measurement of the cohesive force

In this study, we used a centrifugal method (Krupp 1967; Nagaashi et al., 2018) to measure the cohesive force between a particle and a plate. Using this method, we can directly, simultaneously, and statistically measure the cohesive forces of irregularly shaped particles.

All cohesive force measurements were conducted in the atmosphere at a relative humidity of 30%–40%. First, the particles were attached to a glass slide, and their optical images were taken using a microscope. Next, the slide was placed in a tabletop centrifuge (himac CT15E) with its normal in the horizontal plane. The centrifugal force in the horizontal plane was applied for a few minutes to the particles that adhered to the slide. The slide was then gently removed from the centrifuge and its optical microscope images were taken from the same location from where the images were taken before the application of the centrifugal force. The slide was once again placed in the centrifuge and a larger centrifugal force was applied to it. Because the centrifugal acceleration $\alpha$ applied in these measurements was in the range of 20–7000 G (G is the gravitational acceleration), the separation force was applied approximately perpendicular to the slide. The procedure was repeated as $\alpha$ was increased in nine steps of 20, 45, 100, 200, 450, 1000, 2000, 4500, and 7000 G. The particles separated from the slide when the centrifugal forces exceeded the cohesive forces between the particles and the slide. The possible range of the centrifugal acceleration at which the particles separated from the slide was determined using optical microscope images.

The mass $M$ of each particle used to calculate the centrifugal force was estimated using the $a$-axis length of each particle measured by the optical microscope images and particle density listed in Table 1, as described in Section 3.2. The cohesive force between the particle and the slide was then obtained as the geometric mean of the two subsequent centrifugal forces applied to the particle $M\sqrt{\alpha_i \alpha_{i+1}}$, $(\alpha_i < \alpha_{i+1})$, where $\alpha_i$ is the maximum centrifugal acceleration at which the particle remained on the slide, and $\alpha_{i+1}$ is the centrifugal acceleration at which the particle no longer remained on the slide.

The number of particles of the meteorite samples and other samples measured in this study were in the ranges of 104–276 and 145–349, respectively. During a previous study by the authors, a significant fraction of the particles remained on the plate even at the maximum permitted centrifugal acceleration (5000 G), their cohesion could not be determined. In this study, however, we were able to measure the cohesion of at least 94% of the particles of each type of sample. Table 1 shows the measured number and fraction of the particles.

2.4 Measurement of the amount of adsorbed water vapor

We used a high-precision gas and vapor physisorption instrument (BELSORP-max II) to estimate the amount of adsorbed water vapor at the ambient pressure at which the cohesive force measurement was taken. The constant volume method was adopted, in which a change in the gas pressure in the container between before and after gas adsorption on the particles was detected. The amount of gas molecules is estimated based on the equation of state of ideal gas, whereby the amount adsorbed on the particles is determined. Before conducting the measurement, vessels containing sub-cm$^3$ samples were evacuated at 150 °C, which is a nominal pretreatment for silica to expose the surface (e.g., Tarasevich, 2007), for 6 h. The amount of adsorption was measured as the relative pressure, which is the ratio of the equilibrium pressure of adsorption to the saturated vapor pressure, was changed. The measurements were conducted on the Allende (CV3) and NWA 1794 (LL5) samples.

## 3. Results
3.1 Particle shape

The cumulative percentages of the axial ratio and circularity of the two-dimensional projection of each type of particle are shown in Figures 2a and b. Because circularity depends on the spatial resolution of the image, we corrected the bias, based on the fractal dimension of the particles, using the procedure described in Appendix A. Table 2 presents the average, standard deviation, and median of the axial ratio and circularity of the particles. Because particles are likely to settle on the glass slide with the *c*-axis approximately perpendicular to the slide, the projected axial ratio of the particles may be considered as *b*/*a*, a zeroth-order approximation. The axial ratios of the particles except for those of the smaller glass powders and silica sand, were similar to one another and to the *b*/*a* for impact fragments (~0.73); Itokawa particles (0.71 ± 0.13); and Itokawa, Eros, and Ryugu boulders (~0.68, 0.71–0.73, and ~0.68, respectively). However, the circularity of the meteorite particles is 0.74–0.78 at 2000–4000 pixels (the circularity measured at a spatial resolution at which the two-dimensional projected area of the particle corresponds to 2000–4000 pixels), while those of Itokawa rocks and large impact fragments is 0.66–0.74 (median) at 3000 pixels (Appendix A), suggesting that the meteorite particles broken in this study using a mortar and a pestle are rounder than the large impact fragments and Itokawa rocks. Meanwhile, the circularity of the small glass powder and silica sand particles is 0.69–0.72, which is slightly smaller than the meteorite particles, same as with the axial ratio.

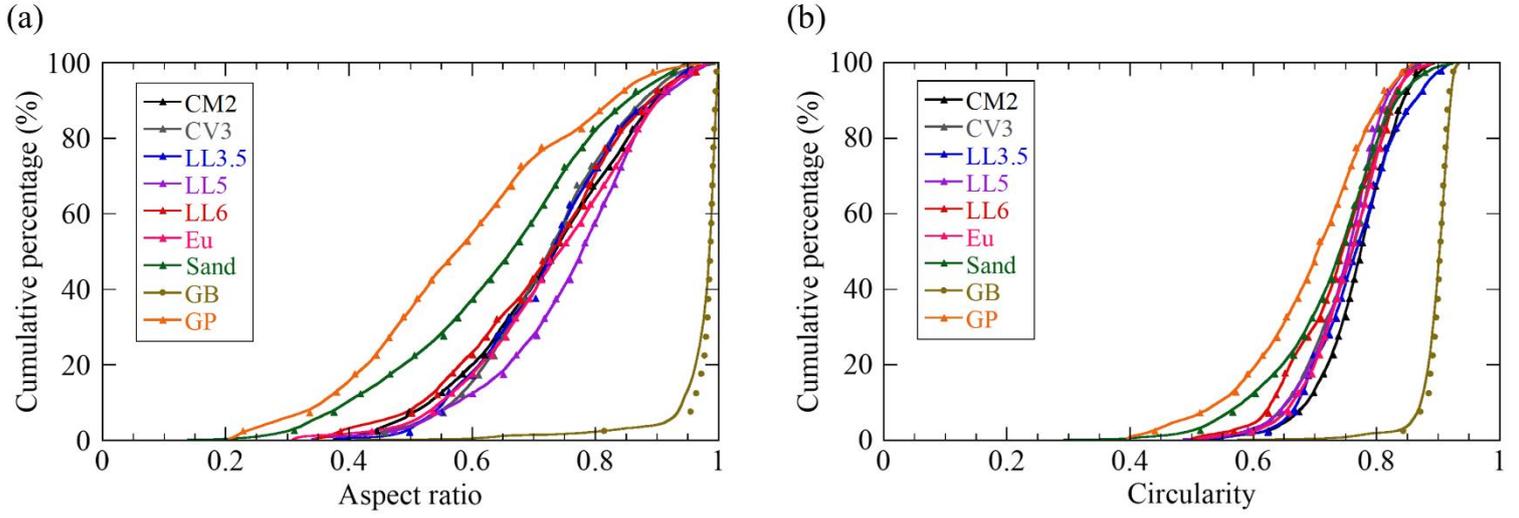

Figure 2. Macroscopic morphology of particles. (a) Axial ratio and (b) circularity of sample particles.

Table 2. Axial ratio and circularity of sample particles.

|  |  | CM2 | CV3 | LL3.5 | LL5 | LL6 | Eu | Sand | GB | GP | Large CM2 | Large GP | Large Sand |
|---|---|---|---|---|---|---|---|---|---|---|---|---|---|
| Axial ratio | Average | 0.72 | 0.72 | 0.72 | 0.76 | 0.71 | 0.73 | 0.64 | 0.97 | 0.58 | 0.72 | 0.73 | 0.70 |
|  | Standard deviation | 0.14 | 0.12 | 0.12 | 0.12 | 0.14 | 0.14 | 0.17 | 0.06 | 0.17 | 0.13 | 0.12 | 0.14 |
|  | Median | 0.74 | 0.73 | 0.73 | 0.78 | 0.73 | 0.74 | 0.67 | 0.99 | 0.58 | 0.73 | 0.75 | 0.71 |
| Circularity | Average | 0.77 | 0.75 | 0.76 | 0.74 | 0.74 | 0.75 | 0.72 | 0.90 | 0.69 | 0.75 | 0.75 | - |
|  | Standard deviation | 0.06 | 0.07 | 0.07 | 0.06 | 0.08 | 0.07 | 0.10 | 0.03 | 0.10 | 0.07 | 0.07 | - |
|  | Median | 0.78 | 0.76 | 0.77 | 0.76 | 0.74 | 0.76 | 0.74 | 0.90 | 0.70 | 0.76 | 0.76 | - |

The smaller glass powder particles and silica sand particles had smaller axial ratios and circularities than those of the other particles used in this experiment, including the larger glass powder and silica sand particles used for the total mass measurement, although in an ideal ellipse, circularity decreases as the axial ratio (*b/a*) decreases. The smaller axial ratio and circularity of the smaller glass powder and silica sand particles may be due to the difference in the positions of the particles, i.e., the larger and smaller particles have similar shapes, but the *c*-axis of the smaller particles is not perpendicular to the glass slide. The glass powder and silica sand particles seem to be composed of flat surfaces, as can be seen in Figure 1g and i. Thus, when they are small and the cohesive force is relatively large, the particles are more likely to settle on the surface of the glass slide with a non-perpendicular *c*-axis orientation. Alternatively, this may be due to a difference in the shape of differently sized particles, i.e., smaller particles do have shorter intermediate axis and less circularity in projected shape.

3.2 Particle mass

The two-dimensional projected image of a particle may correspond to its cross section through the long and intermediate axes of an ellipsoid with an axial ratio typical to impact fragments (Fujiwara et al., 1978). The mass $M$ of the particle was estimated in a previous study (Nagaashi et al., 2018) using the following equation:

$$M = 0.45\rho S^{3/2}, \tag{1a}$$

where $S$ and $\rho$ are the projected area and particle density, respectively. The total $\rho S^{3/2}$ of the 553 large Murchison (CM2) and the 101 large glass powder particles were $\sum \rho S^{3/2} = 8.2 \times 10^{-5}$ kg and $1.1 \times 10^{-4}$ kg, respectively. We assumed the measured (grain) density of the large Murchison for $\rho$. However, the total $M$ of the 553 large Murchison particles and the 101 large glass powder particles were $\sum M = 7.7 \times 10^{-5}$ kg and $5.6 \times 10^{-5}$ kg, respectively. The results suggest that the estimation made on the basis of an ellipsoidal (with $a{:}b{:}c = 2{:}\sqrt{2}{:}1$) assumption overestimates the masses of the glass powder particles by a factor of ~2. In contrast, the ellipsoidal assumption predicts the masses of the meteorite particles, which are 1.1 times the measured masses. If we assume that the bulk density of Murchison is 2.31 g/cm$^3$, as given in the literature (Macke et al., 2011b), the $\sum \rho S^{3/2}$ of the large Murchison particles is $6.4 \times 10^{-5}$ kg. The estimation based on the ellipsoidal assumption is 0.83 times the measured mass.

The *c*-axis of a small-sized particle may significantly incline from the line of sight. To avoid this possible effect on the projected area, we estimated the mass of the particles using the long diameter of the projected plane, the *a*-axis, rather than the projected area *S*, although the assumption of the previous study (Nagaashi et al., 2018) may produce an appropriate estimation for meteorite particles. We estimated the mass of small particles using the following equation:

$$M = 0.52\rho a^3 \text{ (glass beads)}, \tag{1b}$$
$$M = 0.08\rho a^3 \text{ (glass powders)}, \tag{1c}$$
$$M = 0.12\rho a^3 \text{ (meteorite particles)}, \tag{1d}$$
$$M = 0.10\rho a^3 \text{ (silica sands)}. \tag{1e}$$

Here, the value for glass beads is based on an assumption for an ideal sphere with a diameter of $a$. For meteorite particles and glass powders, we assumed that the large particles measured their mass and the small particles measured their cohesive force had similar axial ratios. We used $\sum \rho a^3 = 6.2 \times 10^{-4}$ kg and $7.1 \times 10^{-4}$ kg and $\sum M = 7.7 \times 10^{-5}$ kg and $5.6 \times 10^{-5}$ kg for the 553 large Murchison particles and the 101 large glass powder particles to obtain $\sum M / \sum \rho a^3 = 0.12$ and $0.08$, respectively. Because the axial ratio of the smaller silica sand particles was between the axial ratio of the small glass powder and meteorite particles, we assumed the average of the latter two axial ratios as the axial ratio of the smaller silica sand particles.

3.3 Surface topography

The surface morphologies of the particles of each sample obtained using a confocal laser scanning microscope are depicted in Figure 3a, and their one-dimensional profiles extracted along a line perpendicular to the line of sight are depicted in Figure 3b. The cumulative percentage, $f$, of the absolute value of the difference between adjacent data points, $\delta h$, is shown in Figure 3c. The data are fitted by a Weibull distribution using three parameters as indicated below:

$$f_{\delta h}(<\delta h) = 1 - \exp\left\{-\left(\frac{\delta h - \delta h_0}{\overline{\delta h}}\right)^{\phi_{\delta h}}\right\}, \qquad (2)$$

where $\overline{\delta h}$ is a typical $\delta h$, $\phi_{\delta h}$ is a parameter that characterizes the width of the distribution, with larger values indicating narrow widths, and $\delta h_0$ is a threshold parameter. For glass beads, glass powders, and silica sand, we set $\delta h_0 = 0$ because it is so small that it is not sufficiently constrained by the measurements taken at a resolution of 0.01 µm. The values obtained for each parameter are summarized in Table 3. The obtained $\overline{\delta h}$ and $\phi_{\delta h}$ for each particle type are shown in Figure 3d. Figures 3c and d indicate that CM2 and LL3.5 particles have a narrower distribution of roughness and a larger fraction of $\delta h$ in the range of 0.1–1 µm than the other types of meteorite particles, indicating the presence of a large number of asperities in the 0.1–1 µm range, which may be related to the degree of thermal metamorphism (Huss et al., 2006). In addition, Figures 3a and b reveal that the surfaces of the glass powder, glass beads, and silica sand particles are much smoother than those of the meteorite fragments, 0.1–1 microns in size.

The surface roughness of the glass slide, characterized using the arithmetic mean roughness $R_a$ (Appendix B), is ~4 nm. The $R_a$ values for meteorite and non-meteorite particles are ~300 nm and ~30 nm, respectively. Table 4 summarizes the mean value and standard deviation of $R_a$ for each particle type.

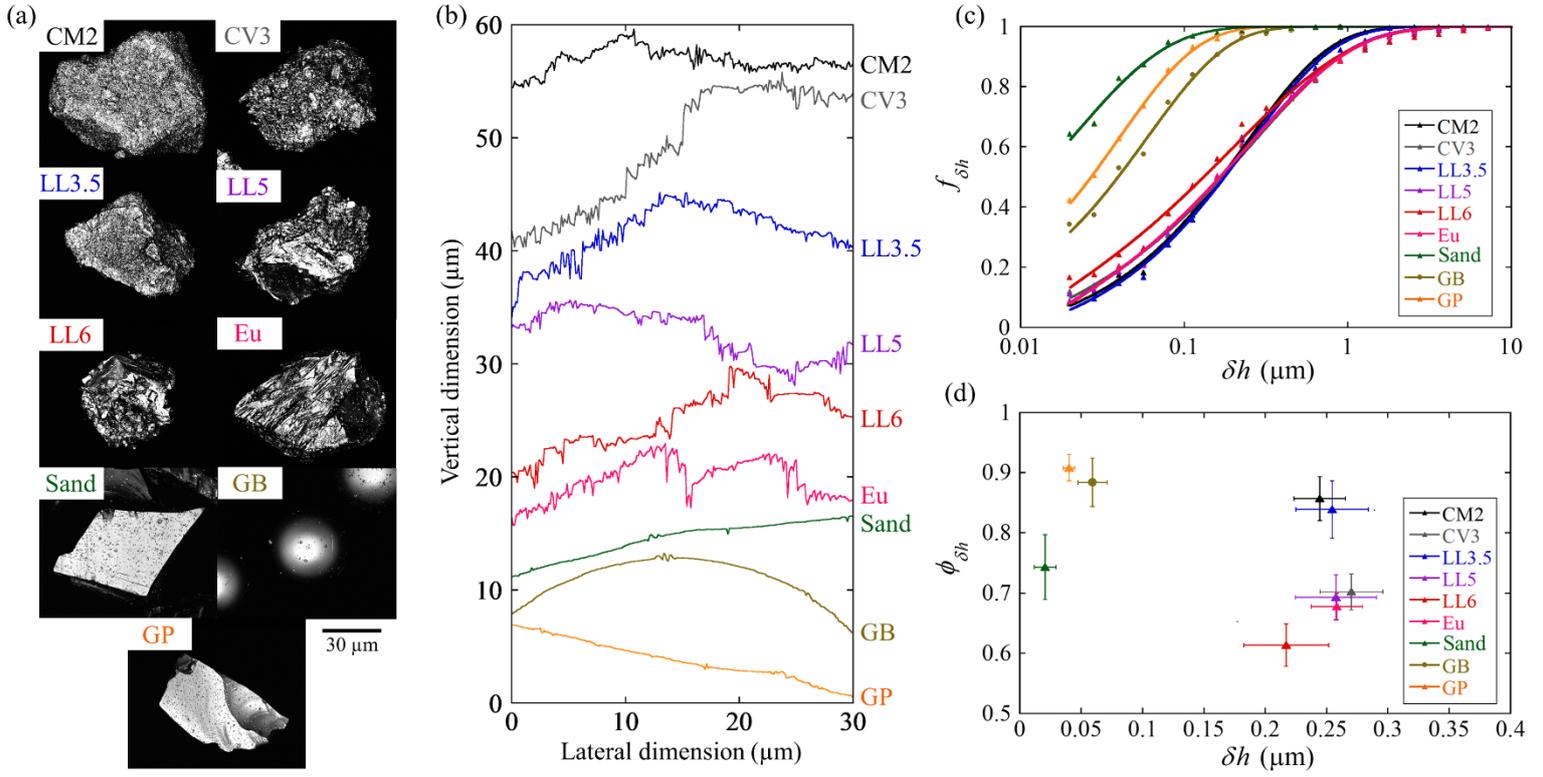

Figure 3. Microscopic morphology of the particle. (a) Surface morphologies of each type of sample particle acquired via confocal laser scanning microscopy and (b) the one-dimensional profile extracted from the data. The horizontal axis shows the location along a line perpendicular to the line of sight, while the vertical axis shows the height. (c) Cumulative percentage of the absolute value of the difference between adjacent data points, $\delta h$, on a profile shown in (b) for each type of sample particle. Each solid curve is a result fitted by Eq. 2. (d) Obtained fitting parameters in (c), $\delta h$ and $\phi_{\delta h}$.

Table 3. Results for the values of $\overline{\delta h}$, $\phi_{\delta h}$, and $\delta h_0$.

|  | CM2 | CV3 | LL3.5 | LL5 | LL6 | Eu | Sand | GB | GP |
|---|---|---|---|---|---|---|---|---|---|
| $\overline{\delta h}$ (μm) | 0.24 ±0.02 | 0.27 ±0.02 | 0.25 ±0.03 | 0.26 ±0.03 | 0.22 ±0.03 | 0.26 ±0.02 | 0.021 ±0.008 | 0.059 ±0.011 | 0.040 ±0.004 |
| $\phi_{\delta h}$ | 0.86 ±0.04 | 0.70 ±0.03 | 0.84 ±0.05 | 0.69 ±0.04 | 0.61 ±0.04 | 0.68 ±0.02 | 0.74 ±0.05 | 0.88 ±0.04 | 0.91 ±0.02 |
| $\delta h_0$ (μm) | 0.0078 | 0.0092 | 0.011 | 0.012 | 0.011 | 0.013 | 0 | 0 | 0 |

Table 4. $R_a$ values of the samples and the glass slide.

|  | CM2 | CV3 | LL3.5 | LL5 | LL6 | Eu | Sand | GB | GP | Slide |
|---|---|---|---|---|---|---|---|---|---|---|
| Average (nm) | 270 | 360 | 290 | 360 | 380 | 390 | 25 | 24 | 32 | 3.9 |
| Standard deviation (nm) | 30 | 100 | 50 | 80 | 120 | 40 | 10 | 8 | 13 | 0.1 |

3.4 Cohesion

Figure 4 shows the cumulative percentage of the measured cohesive force. For comparison, a model distribution of cohesive forces, based on the JKR theory (Johnson et al., 1971), is also shown in Figure 4. Here, the model distributions are those expected by the theory in the case of silica spheres with a size distribution of carbonaceous chondrite particles. As the size distribution for the model, a synthesized distribution composed of all the Murchison and Allende particles was used in this study. In the JKR theory, the force $F_{\text{theo}}$ required to separate two spheres in contact is given as follows:

$$F_{\text{theo}} = 3\pi\gamma R, \quad (3)$$

where $\gamma$ is the surface energy of the particles and $R$ is the reduced radius of the two spheres of radii $R_1$ and $R_2$, respectively, which is given by the following equation:

$$\frac{1}{R} = \frac{1}{R_1} + \frac{1}{R_2}. \quad (4)$$

Because the contact is between a glass slide and a particle, $R$ is equal to the particle radius. We used the surface energy of silica particles $\gamma = 0.025$ J/m$^2$, which was obtained through a measurement in the atmosphere (Kendall et al., 1987), for both the particles and glass slide. In the calculation of the model cohesive forces, we used the equivalent circle radius as the radius $R$ of each particle. The particle masses were estimated using Equations 1b-1e and the particle densities listed in Table 1.

The cohesive force measurements are fitted with a Weibull distribution with three parameters, similar to Equation 2, and are written as follows:

$$f_{F_{\text{meas}}}(< F_{\text{meas}}) = 1 - \exp\left\{-\left(\frac{F_{\text{meas}} - F_{\text{meas0}}}{\overline{F_{\text{meas}}}}\right)^{\phi_{F_{\text{meas}}}}\right\}, \quad (5)$$

where $\overline{F_{\text{meas}}}$ is a typical $F_{\text{meas}}$, $\phi_{F_{\text{meas}}}$ is a parameter that characterizes the width of the distribution, and $F_{\text{meas0}}$ is a threshold parameter. The values obtained for each parameter are summarized in Table 5.

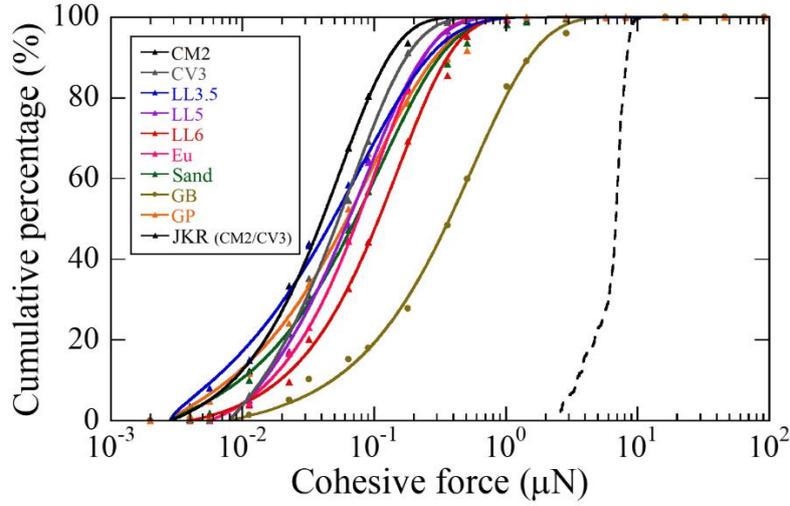

Figure 4. Cumulative percentage of the experimentally measured values of cohesive force. Dashed curve is a model distributions of cohesive forces, predicted by the JKR theory in the case of silica spheres with a size distribution of carbonaceous chondrite particles used in this study.

Table 5. Results for the values of $\overline{F_{\text{meas}}}$, $\phi_{F_{\text{meas}}}$, and $F_{\text{meas0}}$.

|  | CM2 | CV3 | LL3.5 | LL5 | LL6 | Eu | Sand | GB | GP |
|---|---|---|---|---|---|---|---|---|---|
| $\overline{F_{\text{meas}}}$ (μN) | 0.055 | 0.068 | 0.078 | 0.087 | 0.15 | 0.10 | 0.11 | 0.56 | 0.097 |
| $\phi_{F_{\text{meas}}}$ | 0.99 | 0.93 | 0.70 | 0.87 | 0.98 | 1.0 | 0.81 | 0.85 | 0.76 |
| $F_{\text{meas0}}$ (μN) | 0.0028 | 0.0080 | 0.0028 | 0.0080 | 0.0040 | 0.0057 | 0.0028 | 0.0080 | 0.0028 |

Figure 4 shows that the range of the measured cohesive forces of the CM2 and the CV3 particles is approximately two orders of magnitude wider than the range of the model cohesive forces estimated by assuming a perfect sphere with a synthesized size distribution. A similar tendency is observed in the other irregular particles. This suggests that the particles of several tens of micrometers in size have a cohesive force whose distribution and values are approximately two orders of magnitude wider and more than one order of magnitude smaller than that of particles estimated on the basis of the perfect sphere assumption, respectively. These results are consistent with the results of a previous study by Nagaashi et al. (2018). The irregular particles have cohesive forces, which are several times lower than those of spherical particles. Among the irregular particles, CM2 particles have slightly lower cohesive forces whereas LL6 particles have slightly higher cohesive forces. If the microporosity of the particles of carbonaceous chondrites are considered, i.e., if a smaller density is assumed, the tendency will be more prominent.

3.5 Water vapor adsorption isotherm

The water vapor adsorption isotherms of CV3 and LL5 measured using a high-precision gas and vapor physisorption instrument described in Section 2.4 are depicted in Figure 5a. The curves agree with the general adsorption isotherms, such as Type II and IV of the adsorption isotherm classification (Sing et al., 1985). The initial bending of the curve as the relative pressure of the water vapor is increased indicates that the surface of the particles is completely covered with a monolayer, and that multilayer adsorption has begun. In this study, the amount of water vapor adsorbed by the time the monolayer was completely formed was determined based on the excess surface work theory (Adolphs and Setzer, 1996), which is a descriptive theory of adsorption isotherms based on thermodynamics. In this theory, the ratio of the change in the chemical potential of water vapor during the isothermal adsorption, $\Delta\mu$, to its change in the chemical potential at the beginning of adsorption $\Delta\mu_0$ can be written using the volume of water vapor adsorption per gram of sample particles at each relative pressure of water vapor, $n_{\text{ads}}$, and the volume of water vapor required to completely cover the surface of one gram of sample particles with a monolayer, $n_{\text{mono}}$, as indicated below.

$$\frac{\Delta\mu}{\Delta\mu_0} = \exp\left(-\frac{n_{\text{ads}}}{n_{\text{mono}}}\right). \tag{6}$$

The chemical potential is expressed as $\Delta\mu = RT \ln(p/p_s)$, where $R$ is the gas constant, $T$ is the temperature, $p$ is the partial water vapor pressure, and $p_s$ is the saturated water vapor pressure. Equation 6 can be transformed as follows:

$$n_{\text{ads}} = -n_{\text{mono}}\ln\left|\ln\left(\frac{p}{p_s}\right)\right| + n_{\text{mono}} \ln\left|\frac{\Delta\mu_0}{RT}\right|. \tag{7}$$

The obtained water vapor adsorption isotherms (Figure 5a) with the natural logarithm of the absolute value of the natural logarithm of relative pressure plotted on the horizontal axis is shown in Figure 5b. According to Equation 7, the slope of the fitting curve of the plot is $-n_{\text{mono}}$. We refrained from using the data for $p/p_s > 0.6$ for fitting because the nature of adsorption for $p/p_s < 0.6$ and $p/p_s > 0.6$ in Figure 5a will be different (Sing et al., 1985). Accordingly, the estimated numbers of particles on the surface, $n_{\text{mono}}$, were 0.58–0.59 cm³/g and 1.23–1.25 cm³/g for CV3 and LL5, respectively. From Figure 5a, the amounts of adsorbed water vapor in the CV3 and LL5 particles at a relative humidity in the range of 30%–40%, $n_{\text{ads}}(p/p_s = 0.3\text{-}0.4)$, are estimated to be 1.11–1.31 cm³/g and 2.71–3.04 cm³/g, respectively. Because the $n_{\text{ads}}(p/p_s = 0.3\text{-}0.4)/n_{\text{mono}}$ of the CV3 and LL5 particles are 1.9–2.3 and 2.2–2.5, respectively, approximately two water-vapor adsorption layers were estimated to be formed on the surface of the CV3 and LL5 particles during the measurements of the cohesive forces of the particles.

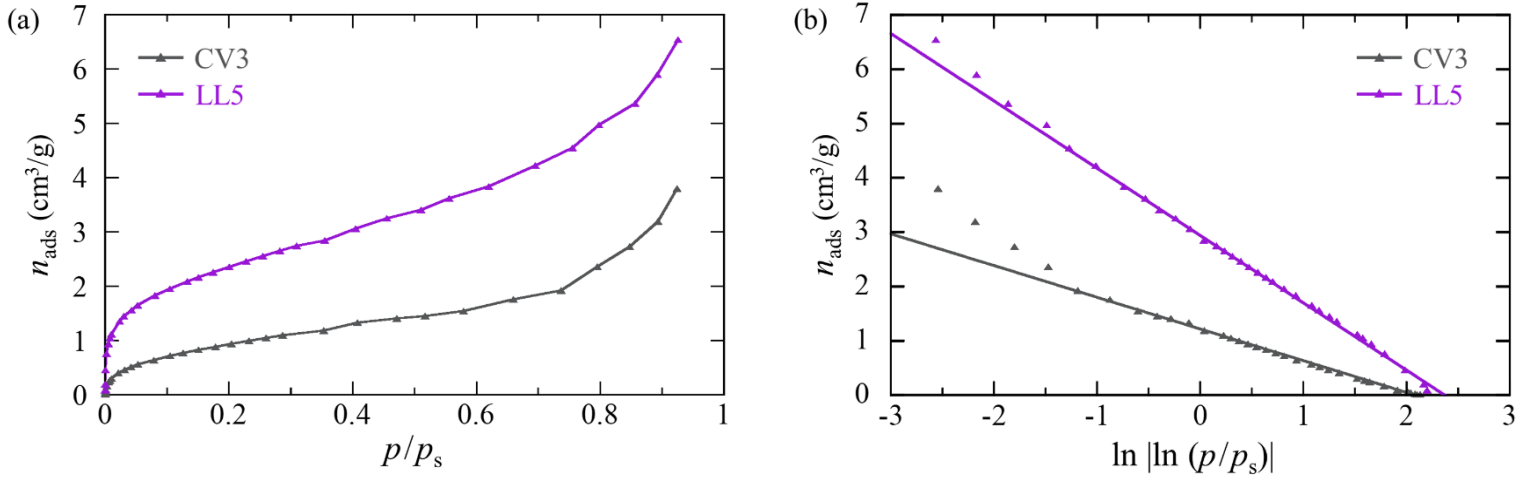

Figure 5. Amount of water vapor adsorption onto particle surfaces. (a) Water vapor adsorption isotherms. (b) Plotted (a) on the horizontal axis as the natural logarithm of the absolute value of the natural logarithm of relative pressure. The slopes of the solid fitting lines represent $-n_{\text{mono}}$.

## 4. Discussion

4.1. Effect of particle surface asperities on the cohesive forces

In Figure 6a, we plot the typical cohesive force shown in Table 5, using $\overline{F_{meas}}$ normalized by the JKR theory prediction for spherical particles, against the mean circularity of the particles. For the JKR theory prediction we considered the surface energy of silica in open air to be 0.025 J/m$^2$ (Kendall et al., 1987). The cohesive forces between spherical glass beads (circularity of 0.933) and irregularly shaped crushed glass particles (circularity of 0.718), ~40 μm in size, and glass substrates of different surface roughness values measured by the impact separation method (Iida et al., 1993) are also shown in Figure 6a. In the previous study (Iida et al., 1993), for a smooth substrate with $R_a$ of a few nm, spherical glass beads exhibited cohesive forces similar to those predicted by the JKR theory, whereas irregularly shaped crushed glass showed a ~4 times lower cohesive force. This decrease of the cohesive force with decreasing circularity of the particles has been generally observed (Iida et al., 1993; Salazar-Banda et al. 2007). We assumed a power-law relationship for the effect of the circularity of the particles on the cohesive force. We also assumed that the spherical glass beads and crushed glass used in the previous study had perfectly smooth surfaces. Using the three data points of the two particles and $(C, \overline{F_{\text{meas}}}/F_{\text{theo}}) = (1, 1)$, we derived the following empirical relationship:

$$\frac{\overline{F_{\text{meas}}}}{F_{\text{theo}}} = C^{5.4 \pm 0.1}. \tag{8}$$

However, because the surface roughness of the particles used in the previous study is unknown, the value of the slope in Equation 8 corresponds to the upper limit value.

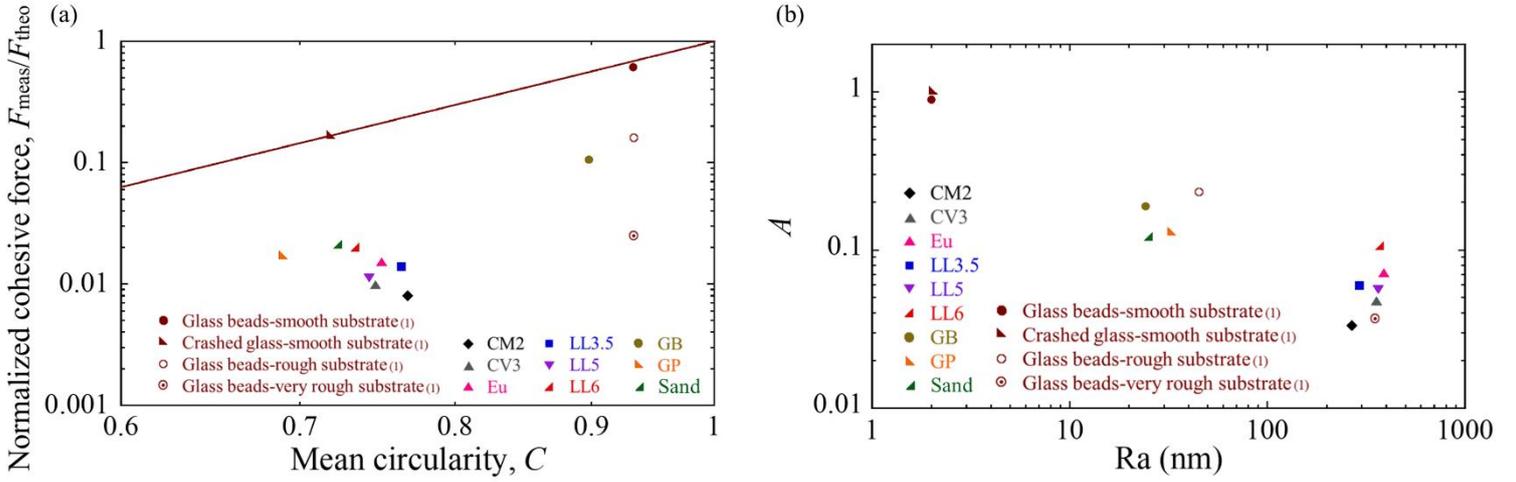

Figure 6. Effects of circularity and surface roughness on cohesive force. (a) Normalized cohesive force plotted against the mean circularity. The solid line is the fitting line of the data of cohesive force of glass beads and crushed glass to smooth substrate in a previous study (Iida et al., 1993; 1). (b) The $A$ in Eq. 9 plotted against the $R_a$ of the particles in this study and of the substrates in Iida et al. (1993).

In the previous study (Iida et al., 1993), spherical glass beads for the rough substrates with $R_a$ of ~40 nm and ~350 nm exhibited ~4 and ~24 times smaller cohesive forces than for the smooth substrates with $R_a$ of a few nanometers. The microscopic surface geometry of the substrates significantly affects the magnitude of the cohesive force. Another study (LaMarche et al., 2017) explained the measured cohesive force of spherical glass beads using two scales of surface roughness rms (nanometers and tens of nanometers) determined for local surface profiles of the particles (15 µm × 15 µm) at one order of magnitude finer resolution than our measurements.

Because the $R_a$ of the glass slide used in this study is ~4 nm, which is comparable to that of the smooth substrate used in the previous study (Iida et al., 1993), the effect of the roughness of the glass slide can be considered negligible. The surface roughness of the spherical glass beads used in this study is ~24 nm, which may explain why the cohesive force of the beads is several times smaller than the force predicted using the JKR theory. The smaller cohesive forces of the irregular particles, that is, the glass powder and silica sand particles, whose $R_a$ are comparable to those of glass beads, could be due to the macroscopic shape effects, as revealed in previous studies (Iida et al., 1993; Salazar-Banda et al., 2007). To characterize the effect of surface roughness $R_a$ (tens of nanometers or more) in addition to circularity $C$, we assumed that the circularity dependence of the cohesive forces of the irregular particles used in this study is represented by the same slope as that of the particles with a perfectly smooth surface. Thus, we assumed the following circularity-dependent empirical relationship for the cohesive force:

$$\frac{\overline{F_{\text{meas}}}}{F_{\text{theo}}} = AC^{5.4 \pm 0.1}. \tag{9}$$

In Table 6, we summarize the values of constant $A$ derived using median $C$ and $\overline{F_{\text{meas}}}$ as well as Equation 3 (the median radius) for each particle type. In Figure 6b, $A$ of each particle type is plotted against its $R_a$. We applied Equation 9 to the data obtained in a previous study by Iida et al. (1993) and plotted the results in Figure 6b, where we used the $R_a$ of the substrates. As the figure shows, $A$ decreases with increasing $R_a$. The meteorite particles have small cohesive forces compared with those of glass powders and silica sand, which can be explained using the size of $R_a$. However, among the meteorite particles, a three-fold difference in $A$ is present between CM2 and LL6. The cohesive forces of CM2 and LL3.5, which have a high percentage of height differences in the range of 0.1–1 μm between adjacent data points in the profiles shown in Figure 3c, are slightly smaller than those of the other meteorite particles. The abundance of surface asperities in this size scale close to the resolution of the confocal microscope may have influenced the cohesive forces.

The empirical relationship in Equation 9 considers the surface roughness of the particles, as in the previous model, which uses two scales of surface roughness to account for the measured cohesive force (LaMarche et al., 2017). However, in the previous model, the small-scale roughness of nanometers dominates the reducing effect on the cohesive force, which differs from the empirical relationship shown in Equation 9, wherein only a roughness exceeding tens of nanometers is considered. Additionally, the cohesive force of 50 μm-sized spherical glass particles measured using atomic force microscope cantilevers (LaMarche et al., 2017) is a fraction of the value determined in this study. This may be attributable to the different measurement methods used. In the case of the cantilever method, the contact between the plate and particles is unstable at one point, but in the case of the centrifugal separation method in this study and the impact separation method (Iida et al., 1993), the contact between the plate and particles can be stable at more than three points.

From the results, the assumption that circularity, $C$ (representing the roughness of up to few tens of μm in this study), and surface roughness $R_a$ (representing the roughness in sub-μm) are responsible for the cohesive forces of the particles can be made. We express the microscopic asperity of the particle, which affects the cohesive force, as $R_{as}$, and the effective (or equivalent) curvature radius, which includes the effects of both $C$ and $R_{as}$, as $R_{eff}$ using the Equations 3 and 9:

$$\overline{F_{meas}} = 3\pi\gamma R_{eff}, \tag{10}$$
$$R_{eff} = R_{as} C^{5.4\pm0.1}, \tag{11}$$
$$R_{as} = AR. \tag{12}$$

The $R_{as}$ derived from Equation 12 is summarized in Table 6 and is typically a few microns. This is approximately the particle size range, in which cohesive forces will be proportional to the particle size (Heim et al. 1999). The circularity of the meteorite particles prepared using a mortar and a pestle is slightly larger than that of either the Itokawa particles or impact fragments ($C \sim 0.7$). Therefore, we use Equation 11 to derive the $R_{eff}$ of each meteorite particle type for a circularity of 0.7. The derived $R_{eff}$ is typically in sub-μm as presented in Table 6.

Table 6. Results for the values of $A$, $R_{as}$, $R_{eff}$, and $F_{regolith}$.

|  | CM2 | CV3 | LL3.5 | LL5 | LL6 | Eu | Sand | GB | GP |
|---|---|---|---|---|---|---|---|---|---|
| $A$ | 0.033 | 0.048 | 0.060 | 0.056 | 0.11 | 0.072 | 0.12 | 0.19 | 0.13 |
| $R_{as}$ (μm) | 0.97 | 1.4 | 1.4 | 1.8 | 3.4 | 2.0 | 2.7 | 4.3 | 2.7 |
| $R_{eff}$ (μm) | 0.14 | 0.20 | 0.21 | 0.27 | 0.50 | 0.29 | - | - | - |
| $F_{regolith}$ (μN) | 0.17 | 0.24 | 0.24 | 0.31 | 0.59 | 0.35 | - | - | - |

In centrifugal methods with preliminary press-on, the cohesive force increases with increasing particle size (Salazar-Banda et al., 2007; Petean & Aguiar, 2015). This increase may be due to the increased contact area resulting from plastic deformation when the force of the press-on exceeds the elastic limit of the surface asperities (Lam & Newton, 1991). According to Hertz theory (Hertz, 1882), when two contacting elastic bodies with curvature radii $R_1$ and $R_2$, Young's moduli $E_1$ and $E_2$, and Poisson's ratios $v_1$ and $v_2$ are compressed against each other by a force $F$, the resulting maximum contact pressure, $P_{\text{max}}$, is given by

$$P_{\text{max}} = \left(\frac{6FE^{*2}}{\pi^3 R^2}\right)^{1/3}, \tag{13}$$

$$\frac{1}{E^*} = \frac{1-v_1^2}{E_1} + \frac{1-v_2^2}{E_2}, \qquad \frac{1}{R} = \frac{1}{R_1} + \frac{1}{R_2}. \tag{14}$$

Using 144 GPa and 0.254 for forsterite as the Young's modulus and Poisson's ratio, respectively (Anderson et al., 1968), and a surface curvature radius of ~300 nm for $R_1$ and $R_2$ and substituting $F = 4/3\,\pi\rho(d/2)^3 g$ (where $d$ is the particle diameter, and $g$ is the gravitational acceleration on the surface of an asteroid) in Equation 13, the equation can be rewritten as follows:

$$P_{\text{max}} \sim 2 \left(\frac{d}{1\text{ cm}}\right)\left(\frac{\rho}{2500\text{ kgm}^{-3}}\right)^{1/3}\left(\frac{g}{10^{-4}\text{ ms}^{-2}}\right)^{1/3}\text{ GPa}. \tag{15}$$

This suggests that the gravitational press-on for 1 cm particles on the Ryugu surface ($g \sim 10^{-4}$ ms$^{-2}$) does not exceed the theoretical strength of silica glass of 24 GPa (Naray-Szabo and Ladik, 1960) and, thus, can be ignored.

4.2. Surface energy in a vacuum

An application of the results obtained in this study to airless environments, such as small bodies, require the consideration of the effect of the measurements performed in the atmosphere. The van der Waals force between two macroscopic spheres is inversely proportional to the square of the inter-surface distance (Israelachvili 2015). The adsorbed water vapor in open air increases the inter-surface distance and makes the cohesive forces, (surface energy), significantly smaller than those that occur with perfect contact in high vacuum environments. The expected surface energy $\gamma_{\text{vac}}$ when particles are in perfect contact in a high vacuum can be written as follows, using the surface energy $\gamma_{\text{air}}$ that is present in open air when there are $n$ layers of water vapor molecules between surfaces (Perko et al., 2001):

$$\gamma_{\text{vac}} = \left(\frac{d_s}{d_s + n\,d_w}\right)^{-2} \gamma_{\text{air}}, \tag{16}$$

where the diameter of the water molecule, $d_w$ is taken as 3.9 Å and the diameter of the SiO$_4$ tetrahedra, $d_s$, is taken as 6.0 Å, based on the length of the O–H bond in H$_2$O (0.96 Å), angle of the H–O–H (104.5°), radius of the H$^+$ ion (1.2 Å), and radius of the O$^{2-}$ ion (1.4 Å) (Fennema, 1996). In this study, it was estimated that ~2 layers of water vapor molecules were adsorbed on the particle surface during the measurements at a relative humidity of 30%–40%. Assuming that ~2–4 layers of water vapor molecules are sandwiched between the surfaces contacting each other, we can estimate that $\gamma_{\text{vac}} = 5$–$13\ \gamma_{\text{air}}$. This estimation is consistent with the estimations of the previous studies, suggesting that the surface energy of silica particles may be 10 times greater in a vacuum than in the atmosphere (Kimura et al., 2015; Steinpilz et al., 2019). Consequently, we use 0.25 J/m² as the surface energy in the following discussion on the cohesive forces in a vacuum.

4.3. Strength of the regolith layers of small bodies

A roughness in the range from sub-micrometers to tens of micrometers is expected to be common in meteorite particles of sizes larger than those used in this study. If this is the case, the effective (or equivalent) curvature radius $R_{\text{eff}}$ is not likely to change with particle sizes larger than those used in this study; that is, the cohesive force would be constant regardless of the particle size. Based on the effects of surface asperities and the differences in the atmosphere and vacuum, the cohesive force of regolith particles can be obtained by substituting $R_{\text{eff}}$ for $R_1$ and $R_2$ in Equations 3 and 4, respectively:

$$F_{\text{regolith}} = 3\pi\gamma_{\text{vac}}\frac{R_{\text{eff}}}{2}. \qquad (17)$$

The result is typically in sub-micronewtons as summarized in Table 6.

The tensile strength of the granular layers on asteroids may be estimated using a relationship used in powder engineering (Rumpf, 1970):

$$\sigma = \frac{9\varphi N F_{\text{regolith}}}{8\pi d^2}, \qquad (18)$$

where $\varphi$ is the filling factor of the regolith layer and $N$ is the coordination number of the regolith particles. Previous studies have reported different relationships between the tensile strength and the cohesive force, and the strength can be scale-dependent (Kimura et al., 2020). Figure 7 shows the relationship between the diameter of a regolith particle and the tensile strength of the regolith layer, which are derived from Equations 17 and 18. We assume that the filling factor $\varphi = 0.5$, the coordination number $N = 6$, and the surface energy $\gamma_{\text{vac}} = 0.25$ J/m² in an airless environment. We used the typical $R_{\text{eff}}$ of CM2 and CV3, and LL3.5, LL5, and LL6 as the range of the $R_{\text{eff}}$ of the regolith particles on C-type and S-type asteroids, respectively. We also show the cases in which the cohesive force is assumed to be proportional to the particle size in Figure 7, as more contact points for larger particles may lead to size dependence.

A breakup event by rotational disruption of C-type, main belt comet P/2013 R3

provides the estimated cohesive strength of the proto-body of 40–210 Pa (Hirabayashi et al., 2014). A fast rotation (1.9529 h) of an S-complex, inner-main belt asteroid (60716) 2000 GD65 requires a cohesion of 150–450 Pa (Polishook et al., 2016). Using the relationship between the tensile strength and particle size shown by Sánchez & Scheeres (2014), the typical particle sizes binding these objects are estimated to be between sub and several microns, while the typical particle sizes estimated on the basis of Figure 7 are several tens of microns. The estimated particle sizes of the asteroids are similar to those of the meteorite particles used in this study; however, the strength model (Equation 18) is a simplified model that does not consider the effect of particle size distribution and the scale effect.

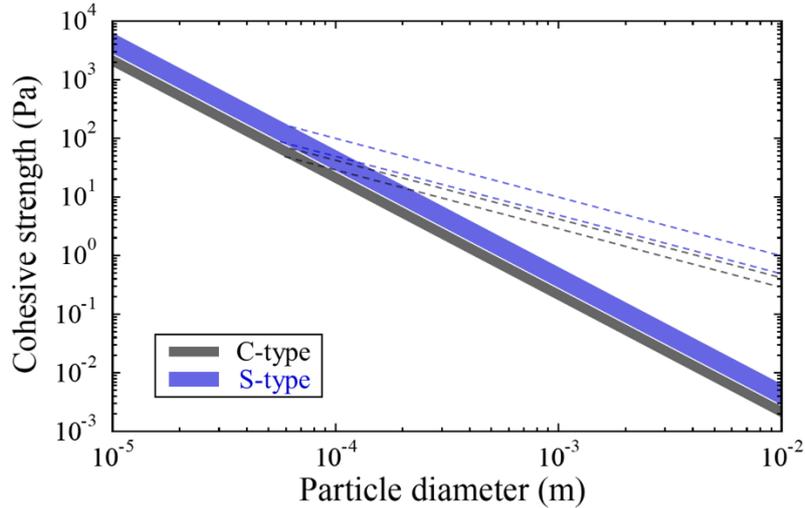

Figure 7. Tensile strength of the regolith layer plotted against the diameter of the regolith particle (Equation 18). The dashed lines ($\sigma \propto d^{-1}$) depict the case in which the cohesive force of larger particles than those used in this study are assumed to be proportional to the particle diameter.

## 5. Summary

We measured the cohesive force of meteorite particles in open air using a centrifugal method and the amount of water vapor adsorbed on the surface under the measurement conditions. The axial ratios and circularities of the particles were measured using optical microscope images, and the surface profiles of the particles were obtained using confocal laser scanning microscopy. We used carbonaceous chondrites (CM2 and CV3), ordinary chondrites (LL3.5, LL5, and LL6), and eucrite particles, prepared by crushing with a median diameter of several tens of microns. For comparison, we also used spherical glass beads, irregularly shaped glass powder particles, and silica sand particles similar in size to meteorite particles.

The axial ratio and circularity of the meteorite particles are almost similar, and the ratio of the *a*-axis to the *b*-axis, and possibly the ratio of *a*-axis to the *c*-axis are similar to those of the impact fragments, Itokawa particles, and rocks although the meteorite particles have a slightly larger circularity than the others. Thus, the meteorite particles used in the study are slightly rounder than those of asteroids. The surface roughness, $R_a$, of the glass slide used in the cohesive force measurements, non-meteorite particles, and meteorite particles are ~4 nm, ~30 nm, and ~300 nm, respectively. In addition, the surface topography of CM2 and LL3.5 were found to exhibit more undulation in the 0.1–1 micron range than other meteorite particles.

The cohesive forces between the glass slide and sample particles had a distribution width of approximately 2 orders of magnitude and 1–2 orders of magnitude smaller than the estimated value of a perfect sphere obtained using the JKR theory. The meteorite particles have smaller cohesive forces than those of other irregular particles. The CM2 and LL3.5 particles have smaller cohesive forces than those of LL6 particles. The difference may be due to the difference in the undulation of the surface topography in the range of 0.1–1 microns. The LL6 particles exhibit a coarser undulation than CM2 and LL3.5 particles, which is probably due to thermal metamorphism.

We demonstrated that the cohesion of meteorite particles is reduced from a perfect sphere by approximately 1/7 owing to the circularity $C$ (representing the roughness of few to tens of µm in this study), and by 1/10–1/25 owing to the fine-scale surface roughness. The effective (equivalent) curvature radius of meteorite particles corresponding to the cohesive force is in the sub-µm range. Because the surface asperities of sub-µm ~ tens of µm in scale are expected to remain unchanged independent of the size of the regolith particles, for particles with sizes above those used in this study, the cohesive force will be constant regardless of the particle size.

We demonstrated that approximately two layers of water vapor were adsorbed to the particle surface under cohesive force measurement conditions. The water vapor layer increases the distance between the particle surfaces and reduces the cohesive force by a factor of up to ~10, which is consistent with the results of previous studies.

Using the measurement results of the cohesive forces and considering the effects of surface asperities and water vapor adsorption, we estimated the cohesive force of particles on asteroid surfaces to be in the sub-µN range. We also estimated the relationship between the tensile strength and the particle size of the regolith layer for C- and S-type asteroids. We demonstrated that the tensile strength of a few hundred pascals in fast-rotating asteroids indicates particles several tens of microns in size.

**Acknowledgments**

We thank S. Dohshi (Osaka Research Institute of Industrial Science and Technology) for the technical assistance provided for measuring the amount of adsorbed water vapor on meteorite particles, T. Okamoto, A. I. Suzuki, and S. Hasegawa (ISAS, JAXA) for providing us with an opportunity to measure the shape of basalt fragments; and K. Yamasaki (Kobe University) for the discussions on the fractal dimensions of fragments. This research was supported by JSPS KAKENHI (No. 18K03723 and No. 19H05081) and the Hypervelocity Impact Facility (formerly the Space Plasma Laboratory), ISAS, JAXA.

## Appendix A

Circularity $C$ is defined as follows:

$$C = \frac{4\pi S}{L^2}, \tag{A1}$$

where $S$ is the projected cross-sectional area of the two-dimensional image of the particle, and $L$ is the perimeter of the two-dimensional image of the particle. We measured the circularity of the rocks on the surface of the asteroid (25143) Itokawa observed using a telescopic optical navigation camera from the home position (HP) and at a low altitude (Fujiwara et al., 2006). We selected 24 unburied rocks shown in the HP images (ST_2489465997, ST_2481442195, ST_2482160259, ST_2485860275, and ST_2506694595 with spatial resolutions of 0.49, 0.41, 0.44, 0.48, and 0.43 m pixel$^{-1}$, respectively) and 61 unburied rocks shown in the close-up images (ST_2539429953, ST_2539437177, ST_2539423137, ST_2539444467, and ST_2539451609 in which the spatial resolutions were 16, 11, 22, 6.0, and 7.8 mm pixel$^{-1}$, respectively). The circularities of the fragments of the impact disruption experiments of basalt targets (68, 41, and 28 fragments in Groups1–3, respectively) were also investigated. Group1 consisted of large fragments from the targets impacted by nylon projectiles 7 mm in diameter with a velocity of approximately 3 km s$^{-1}$ (2.8 km$^{-1}$, unpublished data; 3.3 km$^{-1}$, B-6 of Nakamura, 1993). Group2 consisted of large fragments from the targets impacted by 4.8 mm nylon projectiles at a velocity of approximately 6 km s$^{-1}$ (5.9 and 6.4 km s$^{-1}$, T. Okamoto, private communication). Group3 consisted of the finer counterpart of Group1. Images were taken using a digital camera and an optical microscope.

Figure A1a shows the circularity and equivalent sphere diameter of the Itokawa rocks and impact fragments. The circularity of the rocks and fragments was ~0.7. The perimeters of the particles depend on the pixel resolution of the image, which is the number of pixels constituting the particles. Figure A1b shows the circularity of the rocks and the fragments plotted against the projected area in pixels. The figure indicates that the Itokawa rocks observed on HP images have relatively larger circularities than those of the close-up images because of the lower resolution of the images. To investigate the effect of the image resolution on circularity, we measured the circularity of several rocks and fragments with reduced pixel resolution. Figure A1b shows the examples of circularity measurements for a rock and a fragment with different spatial resolutions. As expected, the higher the resolution (large projected area in pixel), the smaller the circularity. The perimeter of natural objects may be approximated by the power of the square root of the area as indicated below:

$$L \propto S^{\frac{D_f}{2}}, \quad (A2)$$

where $D_f$ is the fractal dimension (e.g., Lovejoy, 1982; Suzuki, et al., 1997). Using the relationship provided in Eq. A2 and Eq. A1, we obtained the following:

$$C \propto S^{1-D_f}. \quad (A3)$$

We averaged the $D_f$ of five randomly selected particles with a larger projected area ($S > 3000$ pixel) for each group and obtained $1.050 \pm 0.016$, $1.024 \pm 0.007$, $1.035 \pm 0.013$, and $1.025 \pm 0.004$ for Itokawa rocks in the close-up images and Group1, Group2, and Group3 fragments, respectively. The circularity of the particles when the projected area of the particle consists of 3000 pixels was estimated using the following equation:

$$C(3000) = C \left(\frac{3000}{S}\right)^{1-D_f}. \quad (A4)$$

Table A1 summarizes the mean and median of the $C(3000)$ values of the rocks and fragments.

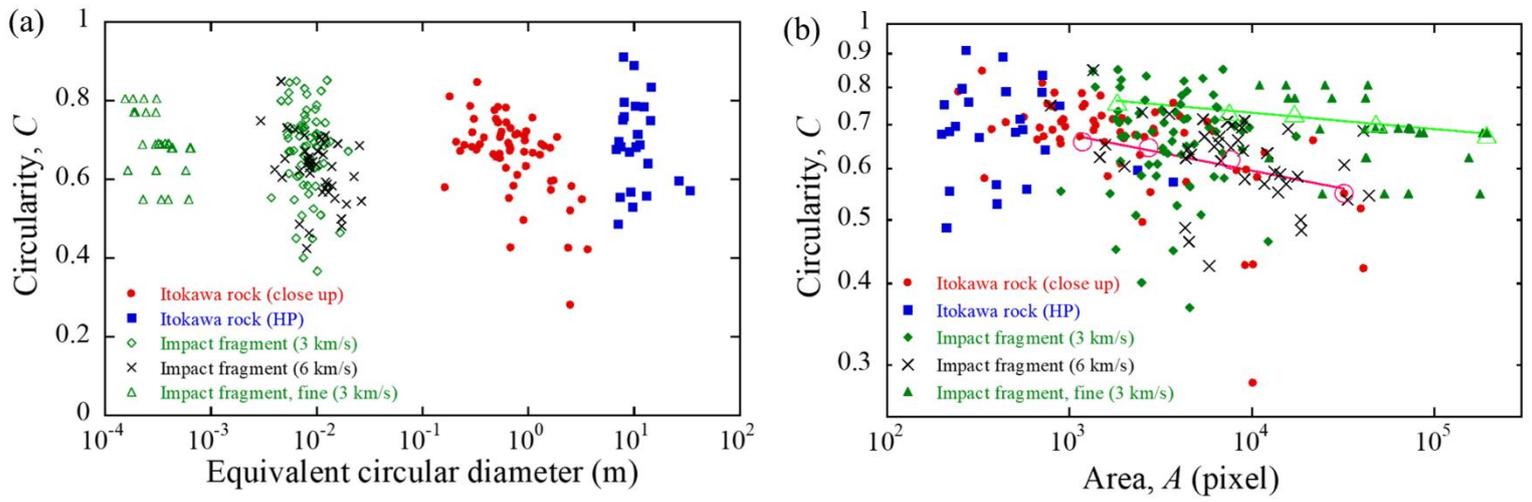

Figure A1. Circularity of Itokawa rocks and impact fragments versus (a) equivalent diameter, and (b) area, in pixel. Circularity values measured at different image resolutions are shown by open circle for a Itokawa rock and open triangle for an impact fragment. The fitted lines according to the power law relationship are also shown.

Table A1. C(3000) values of Itokawa rocks and impact fragments.

| | Itokawa rocks | Impact fragments | | |
| --- | --- | --- | --- | --- |
| | | Group1 | Group2 | Group3 |
| Numbers | 85 | 68 | 42 | 28 |
| C(3000) average | 0.64 ± 0.09 | 0.70 ± 0.07 | 0.64 ± 0.08 | 0.73 ± 0.06 |
| C(3000) median | 0.66 | 0.71 | 0.65 | 0.74 |
| b/a (average) | 0.62 ± 0.16 | 0.68 ± 0.14 | 0.67 ± 0.12 | 0.71 ± 0.14 |

# Appendix B

The arithmetic mean roughness, $R_a$, is defined as follows:

$$R_a = \frac{1}{l}\int_0^l |h(x) - f(x)|\,dx, \tag{B1}$$

where $h(x)$ is the measured height, $f(x)$ is the average curve of the surface, and $l$ is the evaluation length (Gadelmawla et al. 2002). We set the evaluation length, $l$, to 20 μm, and randomly selected three or four one-dimensional profiles (such as shown in Figure 3b) for each particle type. The average curve, $f(x)$, is obtained by $m$-order polynomial fitting to the selected data sets. The values of $m$ are determined by calculating the Akaike information criterion (AIC; Akaike, 1974) given as follows:

$$\text{AIC} = n\left\{\ln\left(\frac{2\pi S_e}{n}\right) + 1\right\} + 2(m+2), \tag{B2}$$

where $n$ is the number of data points and $S_e$ is the residual sum of squares. The lower the AIC value, the better the fit. Figure B1 shows an example of the data set used and the fitting curves.

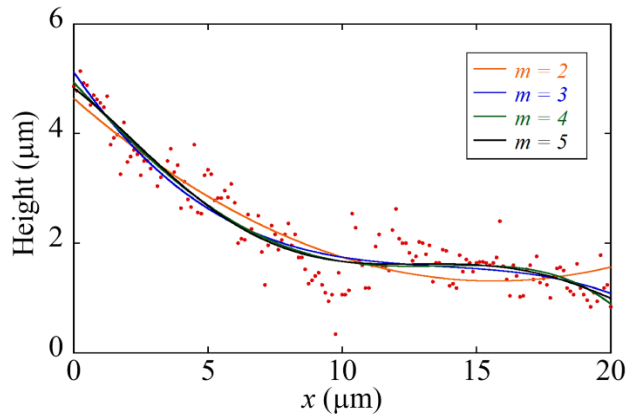

Figure B1. Example of fitting curves for calculating the Ra of particles. In this case, the AIC value became minimal at $m = 4$.